*An In Situ Surface-enhanced Infrared Absorption Spectroscopy Study of Electrochemical CO$_2$ Reduction: Selectivity Dependence on Surface C-Bound and O-bound Reaction Intermediates*

Yu Katayama[1,4], Francesco Nattino[5], Livia Giordano[3], Jonathan Hwang[2], Reshma R. Rao[3], Oliviero Andreussi[6], Nicola Marzari[5], and Yang Shao-Horn[1,2,3]

[1] *Research Laboratory of Electronics,*

[2] *Department of Materials Science and Engineering,*

[3] *Department of Mechanical Engineering, Massachusetts Institute of Technology, Cambridge, MA 02139, USA*

[4] *Department of Applied Chemistry, Graduate School of Sciences and Technology for Innovation, Yamaguchi University, Tokiwadai, Ube, 755-8611, Japan*

[5] *Theory and Simulation of Materials (THEOS) and National Centre for Computational Design and Discovery of Novel Materials (MARVEL), École Polytechnique Fédérale de Lausanne, 1015 Lausanne, Switzerland*

[6] *Department of Physics, University of North Texas, Denton, TX 76207, USA.*

ABSTRACT

The $CO_2$ electro-reduction reaction (CORR) is a promising avenue to convert greenhouse gases into high-value fuels and chemicals, in addition to being an attractive method for storing intermittent renewable energy. Although polycrystalline Cu surfaces have long known to be unique in their capabilities of catalyzing the conversion of $CO_2$ to higher-order C1 and C2 fuels, such as hydrocarbons ($CH_4$, $C_2H_4$ etc.) and alcohols ($CH_3OH$, $C_2H_5OH$), product selectivity remains a challenge. Rational design of more selective catalysts would greatly benefit from a mechanistic understanding of the complex, multi-proton and multi-electron conversion of $CO_2$. In this study, we select three metal catalysts (Pt, Au, Cu) and apply *in situ* surface enhanced infrared absorption spectroscopy (SEIRAS) and ambient-pressure X-ray photoelectron spectroscopy (APXPS), coupled to density-functional theory (DFT) calculations, to get insight into the reaction pathway for the CORR. We present a comprehensive reaction mechanism for the CORR, and show that the preferential reaction pathway can be rationalized in terms of metal-carbon (M-C) and metal-oxygen (M-O) affinity. We show that the final products are determined by the configuration of the initial intermediates, C-bound and O-bound, which can be obtained from $CO_2$ and $(H)CO_3$, respectively. C1 hydrocarbons are produced via $OCH_{3,ad}$ intermediates obtained from O-bound $CO_{3,ad}$ and require a catalyst with relatively high affinity for O-bound intermediates. Additionally, C2 hydrocarbon formation is suggested to result from the C-C coupling between C-bound $CO_{ad}$ and $(H)CO_{ad}$, which requires an optimal affinity for the C-bound species, so that $(H)CO_{ad}$ can be further reduced without poisoning the catalyst surface. It is suggested that the formation of C1 alcohols ($CH_3OH$) is the most challenging process to optimize, as stabilization of the O-bound species would both accelerate the formation of key-intermediates ($OCH_{3,ad}$) but also simultaneously inhibit their desorption from the catalyst surface. Our findings pave the way towards a design strategy for CORR catalysts with improved selectivity, based on this experimental/theoretical reaction mechanisms that have been identified. These results also suggest that designing the electronic structure of the catalyst is not the sole determining factor to achieve highly-selective CORR catalysis; rather, tuning additional experimental reaction conditions such as electrolyte-intermediate interactions becomes also critical.

TOC

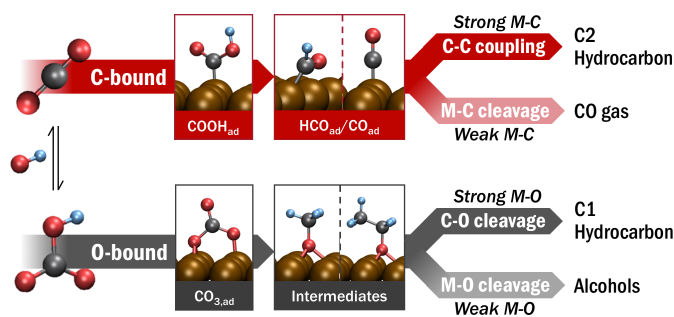

TEXT

1. **Introduction**

The electrochemical reduction of $CO_2$ to fuels and chemicals using renewable energy sources holds great promise for closing the carbon cycle by providing a route to recycle atmospheric $CO_2$[1]. However, poor efficiency and product selectivity remain a major challenge[2,3]. Various metal electrodes have been tested as $CO_2$ reduction catalysts by Hori and coworkers, who broadly categorized the metal surfaces considered into 4 groups, in accordance with the observed product selectivity[4,5]: (1) Pb[6], Hg[7], In[5,7], Sn[6], Cd[5-6], and Tl[8] produce formate as major product (formate forming metals); (2) Au[6,7], Ag[6], Zn[9], Pd[7], and Ga[6,7] form CO as major product (CO forming metals); (3) Cu[9][10][11] and Cu alloys[12][13][14] produce hydrocarbons and alcohols (hydrocarbon forming metals); and (4) Ni[15][16], Fe[16], Pt[17][16], and Ti[7][5] produce hydrogen via the hydrogen evolution reaction (HER) and are inactive for $CO_2$ reduction. Although such classification appears loosely related with their position in the periodic table, the physical origin of their product selectivity remains unknown.

The optimization of the binding energy of O-bound and C-bound adsorbates is known to be critical to control the activity and the selectivity of the CORR[18][19]. However, since the elementary steps involving the adsorption of O-bound species exhibit much smaller overpotentials than those of C-bound intermediates [20], the importance of oxygen-binding energetics is often overlooked. A recent theoretical study by Nørskov and co-workers[21][22] has shown a volcano relationship between the surface binding energy of CO and the associated reaction overpotential as well as the product selectivity, suggesting that $CO_{ad}$ is one of the key intermediates for hydrocarbon-generation from the $CO_2$ reduction reaction (CORR) [23][24]. For strong CO-binding surfaces (left leg of the volcano), such as Pt, the rate-determining step (RDS) is the hydrogenation of $CO_{ad}$ to form $HCO_{ad}$, which explains why the surface is poisoned by $CO_{ad}$ and inactive for the CORR [20][23]. On the other hand, for surfaces that bind CO too weakly (right leg of the volcano), such as Au, the hydrogenation of $CO_2$ to form $COOH_{ad}$ (or, in other words, the adsorption of $CO_2$) is the RDS[20]. Remarkably, Cu, which is known to produce hydrocarbons, is located close to the top of the volcano[3][21][22][23], where the RDS in the formation of hydrocarbons from $CO_2$ is the hydrogenation of $CO_{ad}$ to form $HCO_{ad}$[25]. Experimental studies have supported these computational findings by showing that the electrochemical reduction of HCOOH under similar conditions yields no detectable hydrocarbon products[26][10] while the CO reduction reaction produces $CH_4$ and $C_2H_4$[27][11][28]. Understanding how the M-CO and M-O(H) binding energetics controls the intermediates present on the surface and governs the reaction pathway is critical to establish further insights in the design of CORR catalysts. Unfortunately, little is known about the surface chemistry of the catalysts and of the surface reaction intermediates during the CORR.

While several spectroscopic techniques have been employed recently to identify the reaction intermediates for CORR, including surface-enhanced Raman spectroscopy (SERS)[29,30,31], infrared reflection absorption spectroscopy (IRRAS)[30,32,33], and attenuated total reflection Fourier-transform infrared spectroscopy (ATR-FTIRS)[34], these techniques present severe limitations. SERS is known to amplify signals from the hotspots where the electromagnetic field is particularly intense, making it difficult to extract information that is representative for the entire electrode surface. IRRAS requires a very thin (<100 μm)[35] electrolyte layer between the electrode and the prism. These structural restrictions can affect the catalytic performance of the substrate by limiting the mass transport to the electrode surface[36]. In addition, the spectroscopic analysis is hampered by the large concentration gradients that characterize the electrode interface, especially for gas-evolving reactions such as CORR[36]. Although the use of an attenuated total reflection (ATR) configuration has been shown to overcome these challenges, the low surface sensitivity of this technique limits the time-resolved monitoring of electrochemical reactions[35]. Surface-enhanced infrared absorption spectroscopy (SEIRAS) bypasses these limitations[35]. Furthermore, this technique is characterized by a high surface sensitivity, which has been shown to allow for the detection of CORR reaction intermediates on several metal surfaces, such as Au [37] and Cu[38,36]. However, all SEIRA spectra on the electrochemical interface are differential spectra which always require a reference spectrum recorded at different potentials or solutions. Unfortunately, the comparison of SEIRA studies across metal substrates is limited by the arbitrariness in choosing the reference spectra for background subtraction.

In this work, we explore the origin of the catalytic activity and the selectivity of the $CO_2$ reduction reaction on three metal surfaces (Au, Cu, and Pt) by tracking the surface-bound intermediates using *in situ* surface-enhanced infrared absorption spectroscopy (SEIRAS) and comparing it with density-functional theory (DFT) calculations. In addition, the chemical reactivity between these metal surfaces and $CO_2$ or $H_2O$ is studied with ambient-pressure X-ray photoelectron spectroscopy (APXPS). We find that C-bound intermediates, such as $COOH_{ad}$ and $CO_{ad}$, are dominant on the Au and Pt surfaces, whereas both C-bound and O-bound intermediates are formed on the Cu surface. In particular, O-bound ($CO_{3,ad}$) and C-bound ($(H)CO_{ad}$) intermediates are suggested to be responsible for the production of C1 and C2 hydrocarbons, respectively. Using our combined experimental and theoretical approach, we propose that the configuration of the initial intermediate, C-bound and O-bound, which is determined not only by the M-C and M-O binding affinity but also by the electrolyte composition close to the surface, plays a critical role in determining the reaction pathway and the product selectivity of the CORR. In addition, this study provides insight on how the energetics of key reaction intermediates and the electrolyte composition near the catalyst surface can guide the reaction towards a desired product.

## 2. Experimental and Theoretical Setup

2.1. Electrocatalyst Preparation

For electrochemical measurements, the working electrodes were prepared by sonicating samples of Au (Sigma Aldrich, 0.05 mm thick, 99.9%), Pt (Sigma Aldrich, 0.05 mm thick, 99.99%), and Cu foils (Alfa Aesar Puratronic, 0.05 mm thick, 99.9999%) in deionized (DI) water. The Cu foil was then electropolished in 85% $H_3PO_4$ (Aldrich, 85 wt% in $H_2O$, 99.99% trace metals basis) for 1 min at 3 V and 10 min at ~1.5–2 V (chosen to be higher than the observed oxidation peak from cyclic voltammetry) vs a carbon paper counter electrode. All samples were contacted with Cu wires (Aldrich, 0.64 mm diameter, 99.995%).

For the *in situ* SEIRA measurements, we used a Pt working electrode composed of a thin (*ca.* 50 nm) Pt film deposited on the Si plate (radius 22 mm, thickness 1 mm, Pier optics) by an electroless deposition method[39]. First, the surface of the Si plate was given a hydrophilic treatment by contacting it with a 40% $NH_4F$ solution for a minute. Then palladium seeds were deposited on the base plane with 1% HF–1 mM $PdCl_2$ for 5 min at room temperature. After rinsing with water, platinum electroless deposition was carried out by contacting the Si plates with the Pt plating solution at 50 °C for *ca.* 12 minutes. The Pt plating solution was prepared by mixing LECTROLESS Pt 100 basic solution (30 mL, Electroplating Engineering of Japan Ltd), LECTROLESS Pt 100 reducing solution (0.6 mL), 28% $NH_3$ solution, and ultrapure water. Magnetron sputtering was used to prepare the Au and Cu electrocatalysts. Those metals were sputtered on the electroless deposited Pt layer at a deposition rate of ~0.2 A/s for a thickness of 25 nm, as measured by an electrochemical quartz crystal microbalance.

2.2. Electrochemical measurements

Electroreduction of $CO_2$ was performed in a 2-compartment H-cell, with the Pt mesh counter electrode separated from the catholyte by a Nafion-117 membrane. The Nafion-117 membrane was sequentially conditioned prior to the electrochemical measurements with 0.5 M $H_2SO_4$, 0.5 M $H_2O_2$, and 0.5 M $KClO_4$ for ~1 hour each, and then rinsed with DI water prior to the electrochemical measurements. Both compartments contained 15 mL of $CO_2$-saturated 0.1 M

KHCO$_3$ electrolytes, which was prepared by vigorously bubbling 0.1 M KOH (Millipore Suprapur, >99.995%) with CO$_2$ for at least 20 min, and the pH was confirmed to be 6.8 before use. The headspace of the cathodic chamber was continuously flushed with CO$_2$ into the sample loop of the gas chromatograph (SRI GC), allowing for on-line gas-phase product analysis. Liquid-phase products were analyzed after electrolysis using $^1$H-NMR, using DMSO and phenol as internal standards. A Biologic SP-200 potentiostat was used for all experiments; potentials were measured against an Ag/AgCl reference electrode and converted to the reversible hydrogen electrode (RHE).

2.3. *In situ* SEIRAS measurements

Details of *in situ* SEIRAS were described elsewhere [40,41,42,43]: A Pt-deposited Si plate was pressed against the flat plane of a hemispherical ZnSe prism (radius 22 mm, Pier optics) in order to increase the signal/noise (S/N) ratio in the lower wavenumber region, below 1100 cm$^{-1}$. The prism was mounted in a spectro-electrochemical three-electrode cell with an Ag/AgCl reference electrode and a platinum-wire counter electrode. The SEIRA spectra were obtained with a FT-IR Vertex 70 (Bruker) spectrometer equipped with a MCT detector. The optical path was fully replaced with N$_2$ gas. The measurements were done with 4 cm$^{-1}$ resolution in the 500–3800 cm$^{-1}$ spectral range; 32 scans were averaged. The SEIRA spectra were recorded using a single reflection ATR accessory (Pike Vee-Max II, Pike Technologies) with a Si plate/ZnSe prism at an incident angle of 70 degrees.

The electrolyte solutions were prepared by mixing KOH (Sigma–Aldrich, >85 wt%) with KHCO$_3$ (Sigma–Aldrich, 99.7 wt%) and ultrapure water. Before every experiment, argon was bubbled through the electrolyte for 15 minutes in order to remove air from the solution. CO$_2$ was bubbled through the electrolyte for 2 hours before every experiment to saturate the solution, and during the experiments the $CO_2$ gas was kept flowing above the solution. After deoxygenation of the electrolyte solution, obtained by purging Ar, the prism surface was cleaned by cycling the potential between 0.05 and 0.90 V *vs.* RHE. For electrochemical measurements, linear sweep voltammetry (LSV) was conducted at room temperature by using a HSV-110 (Hokuto Denko) potentiostat. All potentials in this paper were referred to the RHE. The current density expressed was normalized by the geometric surface area (expressed as A cm$^{-2}_{GEO}$). All spectra are shown in absorbance units defined as $\log(I_0/I)$, where $I_0$ and $I$ represent the spectra at the reference and sample potentials, respectively. The reference spectrum $I_0$ was measured at 0.05 V in the blank KOH solution.

2.4. Ambient-pressure XPS measurements

AP-XPS measurements were collected at the Beamline 9.3.2 at Lawrence Berkeley National Laboratory's (LBNL) Advanced Light Source. Films of Au, Cu and Pt ~25 nm thick were sputtered on a Si wafer and placed on a ceramic heater. A thermocouple was then mounted on the sample surface to probe the surface temperature. The films were cleaned by successive Ar$^+$ sputtering followed by annealing in oxygen atmospheres up to 500°C until there was no contribution from the surface carbon or oxygen species.

The C1s spectra have been aligned to the adventitious carbon peak at 284.8 eV. We note that aligning to the metal core levels (Cu 3p, Au 4f and Pt 4f) at the same incident photon energy does not change the corrected binding energy. In order to ensure that the measured spectra are not influenced by radiation damage, spectra were measured at the same spot one hour apart. On a clean sample, $CO_2$ was dosed at room temperature (100 mTorr), followed by 10 mTorr of $H_2O$. The $H_2O$ was prepared from DI water (Millipore, >18.2 M Ω cm). Several freeze-pump-thaw cycles were employed to effectively degas this sample.

At each experimental condition, the C1s spectra was measured at an incident photon energy of 490 eV, and the O1s spectra was measured at an incident photon energy of 735 eV and 690 eV. In addition, the Si 2p core level was also measured at an incident photon energy of 350 eV to ensure no contribution from the substrate through diffusion of Si at high temperatures and/or exposure upon sputtering (Figure S1).

2.5. Computational methods

DFT calculations were carried out using the plane wave Quantum ESPRESSO (QE) distribution [44,45]. We used the PBE generalized-gradient approximation (GGA) [46,47] for the exchange-correlation functional and pseudopotentials[48,49,50] from the Standard Solid-State Pseudopotential library [51,52] (SSSP efficiency 0.7). The energy cutoff for the plane-wave expansion of the wavefunction/electron density was 40 Ry/320 Ry for Cu and Pt and 45 Ry/360 Ry for Au. The first Brillouin zone was sampled with a Γ-centered 6x6x1 k-point grid. A Marzari-Vanderbilt smearing with a conservative width of 0.005 Ry was employed [53], as compatible with the relatively dense k-point sampling. The metal surfaces were modeled by 4-layer-thick metal slabs, which were constructed using the calculated equilibrium value for the bulk lattice constant (3.631 Å for Cu, 3.961 Å for Pt and 4.150 Å for Au). The unit cell was chosen so that periodic replicas are 15 Å apart from each other along the surface normal, and an additional real-space dipole correction has been applied [54]. The three uppermost atomic layers have been relaxed for each of the adsorbates, with the frozen bottommost layer avoiding the non-physical buckling of the metal slabs. Three Cu surfaces (i.e. Cu(100), Cu(111) and Cu(211)) were considered in order to account for the variety of surface terminations that are likely to characterize the polycrystalline electrode surface. The Cu(100) and Cu(111) surfaces were modeled with (3x3) supercells while a (1x3) supercell was used for the Cu(211) surface. For Pt and Au, we have considered only the (111) surface, with a (3x3) supercell.

The presence of the solvent was accounted for by using self-consistent continuum solvation (SCCS) [55], as implemented in the ENVIRON module[56] for QE. Briefly, the slab-adsorbate system was embedded in a smoothly-varying polarizable dielectric medium that mimics the electrostatic response of the solvent. The simulation cell was thus divided in a solute quantum-mechanical region and a solvent region, with dielectric constant $\varepsilon=1$ (vacuum) and $\varepsilon=78.3$ (bulk water), respectively. The boundary between the two regions was defined as a function of the solute's electron density, with parameters $\rho_{min}=0.0001$ a.u. and $\rho_{max}=0.005$ a.u. [55]. Additional non-electrostatic solute-solvent interactions were accounted for through a quantum-surface dependent contribution, with the $\alpha+\gamma$ parameter set to 11.5 dyn/cm (fitg03 of the SCCS parameterization) [55].

Adsorbate vibrational frequencies were computed for the minimum-energy structures as selected from the test of various adsorption sites and configurations. We note in passing that while GGA density functionals typically predict the wrong preferential adsorption sites for CO on transition-metal surfaces [57], they generally return CO-stretching frequencies that are in good agreement with experimental data [58]. Frequencies were calculated through standard finite difference methods, using the relevant tool implemented in the atomic simulation environment (ASE) package [59]. The magnitude of the atomic displacements was set to 0.010 Å, and two (opposite) displacements per atom and cartesian direction were considered to construct the adsorbate's force constant matrix.

SEIRAS-active vibrational modes were identified by estimating the IR cross section[60] of the computed modes from the change in the system's dipole moment along the surface normal.

## 3. Results

3.1. Interaction of $CO_2$ and water with Au, Cu, and Pt

Here we probe the reactivity of the three metal surfaces, i.e. Au (CO-producing metal, weak metal-CO binding), Cu (hydrocarbon-producing metal, intermediate metal-CO binding), and Pt ($H_2$-producing metal, strong metal-CO binding), with $CO_2$ and water by examining the surface reactions with ambient-pressure XPS (APXPS). We find *sp*³ carbon (adventitious carbon), carbonate, and chemisorbed $CO_2$ as major surface species on Au, Cu and Pt, respectively (Figure 1).

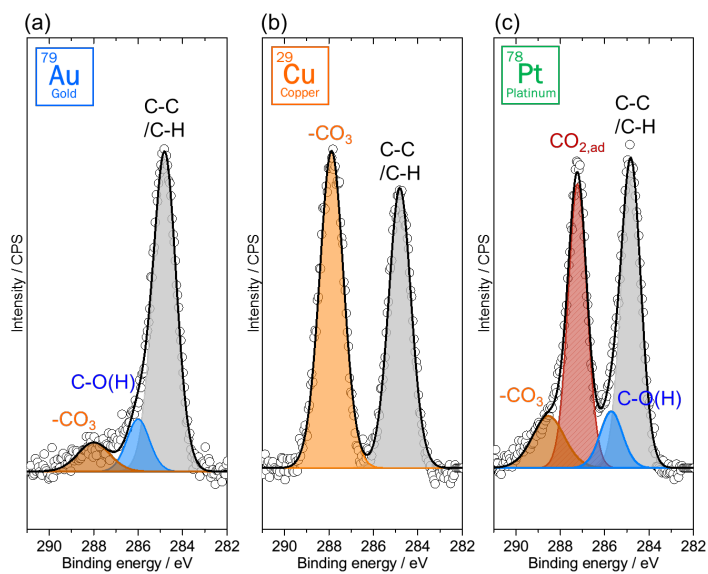

Figure 1 APXPS spectra of the C1s photoemission lines for (a) Au, (b) Cu, and (c) Pt surfaces collected under $p(CO_2)$ = 100 mTorr and $p(H_2O)$ = 10 mTorr at room temperature. All spectra are calibrated with adventitious hydrocarbons at 284.8 eV and background corrected using a Shirley background. C1s spectra are assigned with the following contributions: C-H/C-C ($E_b$=284.8 eV) [61][62][63], C-O(H) ($E_b$~285.5 eV) [61][64][65], $CO_{2,ad}$ ($E_b$~286.8 eV) [61], and $CO_3$ ($E_b$~288.1 eV) [61].

The APXPS measurements were performed under 100 mTorr of $CO_2$ and 10 mTorr of $H_2O$ at room temperature, after the surface was cleaned by repeated cycles of $Ar^+$ sputtering and annealing at 500 °C under UHV conditions (Figure S2). The C1s spectrum for the Au surface showed an asymmetric band, which could be deconvoluted into three peaks that we assigned to $sp^3$ carbon (C-C/C-H, 284.8 eV) [61], carbon with C-O(H) bonds (~286.1 eV) [61], and carbonate (-$CO_3$, ~288.1 eV) [61] (Figure 1a). For the Cu surface, two distinct symmetric bands were observed in the C1s region (Figure 1b), which could be assigned to $sp^3$ carbon (284.8 eV) [61][62][63] and carbonate (~288.1 eV) [61][62]. The spectrum of the Pt surface could be deconvoluted into four peaks that can be assigned to $sp^3$ carbon (284.8 eV) [61], carbon with C-O(H) bonds (~285.8 eV) [61][64][65], chemisorbed $CO_2$ ($CO_{2,ad}$, 287.2 eV) [61], and carbonate (~288.5 eV) [61]. Such speciation is consistent with the O 1s core-level spectra measured for the three metal surfaces under the same conditions (Figure S3). After annealing under ultra-high vacuum (UHV), we observe near-surface oxygen ($O_{int}$) in the low-binding energy region of the AP-XPS O 1s at 529.6 eV, which could originate from either the OH or the residual molecular oxygen in the chamber (~submonolayer oxygen). The quantity of $O_{int}$ species observed in the AP-XPS increases as Au < Pt < Cu, consistently with the M-O binding-energy trends for these three metals.[66] Similarly, the quantity of carbonate ($CO_3$) observed in the C1s spectra (Figure 1) followed the same trend, indicating that the formation of carbonate and $O_{int}$ is somehow related. Potentially, residual OH and $O_2$ in the chamber serve as a source for $O_{int}$, which proceeds to react with $CO_2$ to form adsorbed carbonate species. This process rationalizes why the Cu surface with the strongest M-O affinity[20] forms carbonate as the major species. The Pt surface with less strong M-O bonds but the strongest affinity for carbon-bound species[20,23] is able to stabilize $CO_{2,ad}$ via metal-carbon bonding while forming an intermediate amount of carbonate among the three metals studied. In contrast Au, which has the weakest affinity towards oxygenated and carbon-bound species[20,23], forms the least amount of either $CO_{3,ad}$ and $O_{int}$ and suppresses the $CO_{2,ad}$ formation. These trends in the thermodynamic stability towards carbonaceous adsorbates ($CO_{2,ad}$ and $CO_{3,ad}$) formed from $CO_2$ on these metal surfaces, as proposed from the APXPS measurements, were supported largely by *in situ* SEIRAS measurements on corresponding metals in 1 M $KHCO_3$ at room temperature (see following Section).

3.2. Analysis of the CORR intermediates

The product distribution and activity of the CORR on these three surfaces was investigated through *in situ* SEIRAS, on-line gas chromatography, and NMR coupled with DFT calculations. *In situ* SEIRAS was performed simultaneously with linear sweep voltammetry (LSV) scanned from 0.3 to -0.9 $V_{RHE}$ for the three metal surfaces in a 1 M $KHCO_3$ electrolyte (Figures 2-4). In addition, we performed chronoamperometry (CA) at three different potentials (-0.3, -0.6, and -0.9 $V_{RHE}$) for 1 h to allow for product accumulation, after which gas-phase and liquid-phase products were analyzed by on-line gas chromatography and NMR, respectively.

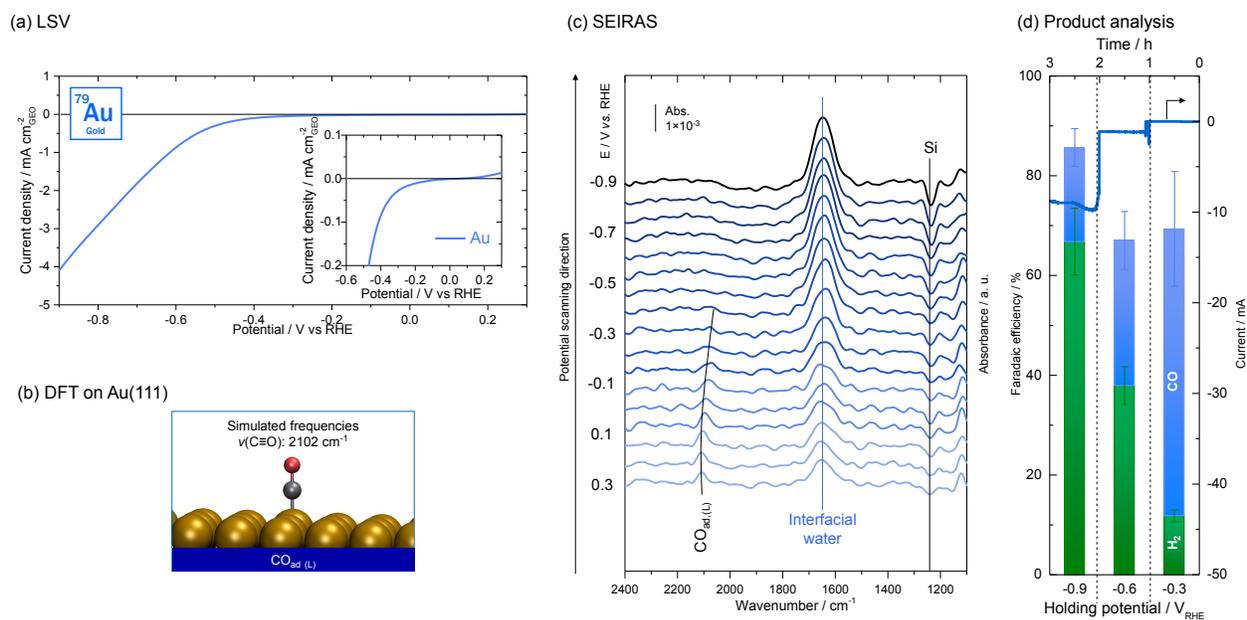

Figure 2 *In situ* ATR-SEIRAS measurement and product analysis on an Au electrode in 1 M $KHCO_3$ (pH=6.8). (a) Linear sweep voltammogram in 1 M $KHCO_3$ at a scan rate of 10 mV s$^{-1}$. Inset shows the magnified linear sweep voltammogram from 0.3 to -0.6 $V_{RHE}$. (b) Calculated infrared-active vibrational frequency of the CO adsorbate on Au(111) and its schematic structure. Au, C, O atoms are depicted as dark-yellow, dark-gray, and red spheres respectively. (c) *In situ* ATR-SEIRA spectra obtained during linear sweep voltammetry in a potential window from 0.3 $V_{RHE}$ to -0.9 $V_{RHE}$ in 1 M $KHCO_3$. A reference spectrum obtained at 0.05 $V_{RHE}$ in 1 M KOH is subtracted. (c) Chronoamperogram and calculated faradaic efficiencies for CORR products after 1 h potentiostatic electrolysis at -0.3, -0.6, and -0.9 $V_{RHE}$.

Selective CO gas formation was found for Au, which is expected to have the lowest affinity for carbon-containing species[23] and oxygenated species[20] (Figure 2). *In situ* SEIRA spectra obtained at 0.3 $V_{RHE}$ showed a band at *ca.* 2100 cm$^{-1}$, which could be assigned to the C≡O stretching mode of linear-bonded CO adsorbates ($CO_{ad,L}$) (Figure 2b) [67][68][69]. The calculated frequency for the C≡O stretching of $CO_{ad,L}$ on Au(111) in implicit water is 2102 cm$^{-1}$, which further supports such assignment (Figures 2b and S7). The $CO_{ad,L}$ band was found to grow from 0.3 to 0.1 $V_{RHE}$, and to subsequently decrease at potentials lower than -0.05 $V_{RHE}$. Below -0.05 $V_{RHE}$ the band was also found to shift towards lower wavenumbers (Figure 2c). Both the redshift and the reduction in intensity of the band can be explained in terms of surface coverage reduction of $CO_{ad,L}$[70][71], indicating the beginning of $CO_{ad,L}$ desorption in this potential range. This hypothesis is consistent with -0.05 $V_{RHE}$ being the onset of the reduction current of $CO_2$ as determined through LSV (Figure 2a, inset). The desorption of $CO_{ad,L}$ from Au is probably due to the displacement of $CO_{ad,L}$ by adsorbed hydrogen atoms ($CO_{ad,L}$ + $H_2O$ + $e^-$ → CO(gas) + $H_{ad}$ + $OH^-$) and/or adsorbed cations (in our case, $K^+$, $CO_{ad,L}$ + $K^+$ + $e^-$ → CO(gas) + $K_{ad}$)[69][72]. The absence of $H_{ad}$-related peaks (*ca.* 2050 cm$^{-1}$) [73][74] could be attributed to the rapid kinetics of the hydrogen-evolution reaction (HER) and the lack of cation-related bands at lower frequencies (below 1000 cm$^{-1}$) [69] could result from the strong IR absorption of the Si ATR crystal in this region. The $CO_{ad,L}$ band at *ca.* 2100 cm$^{-1}$ completely disappeared at potentials lower than -0.3 $V_{RHE}$, at which CO gas was detected as the major products by the on-line GC measurement (Figure 2d). The formation of CO gas persisted at -0.6 and -0.9 $V_{RHE}$ while the $CO_{ad,L}$ band at *ca.* 2100 cm$^{-1}$ was absent in the SEIRA spectra, indicating rapid CO-formation kinetics at large overpotentials (below -0.3 $V_{RHE}$). The amount of CO gas diminishes with decreasing potential and $H_2$ became the major product below -0.6 $V_{RHE}$.

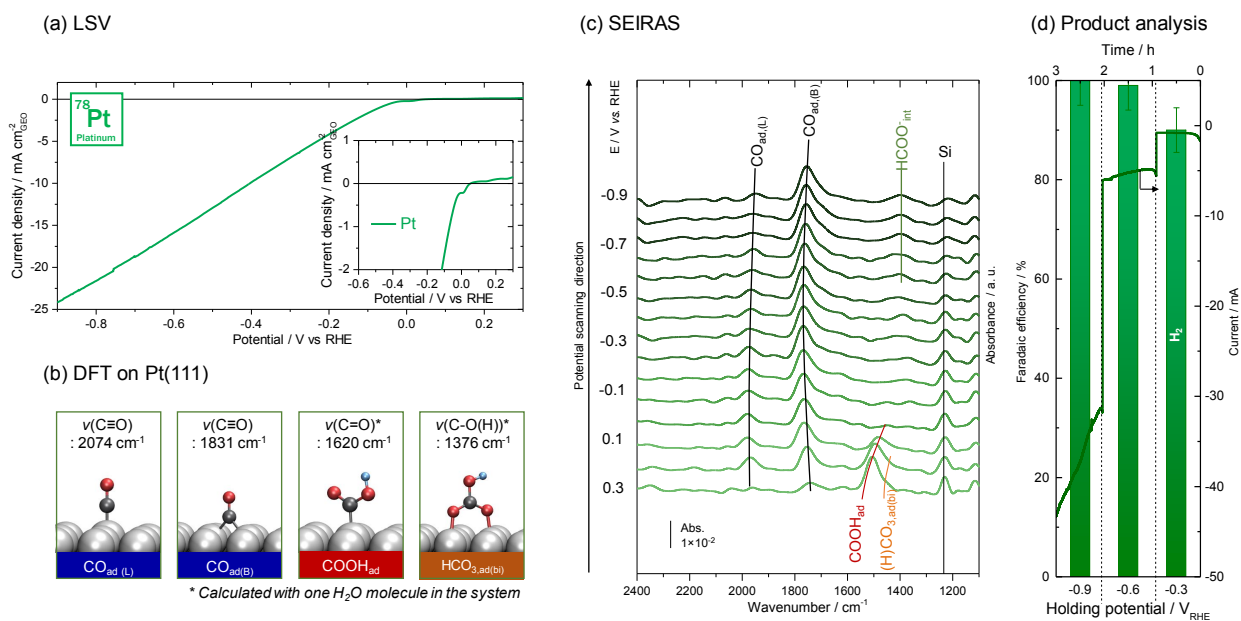

Figure 3 *In situ* ATR-SEIRAS measurement and product analysis on a Pt electrode in 1 M KHCO$_3$ (pH=6.8). (a) Linear sweep voltammogram in 1 M KHCO$_3$ at a scan rate of 10 mV s$^{-1}$. Inset shows the magnified linear sweep voltammogram from 0.3 to -0.6 V$_{RHE}$. (b) Calculated infrared-active vibrational frequencies of possible adsorbed intermediates on Pt(111) and their schematic structures. Pt, C, O, and H atoms are depicted as gray, dark-gray, red and light blue spheres respectively. (c) *In situ* ATR-SEIRA spectra obtained during linear sweep voltammetry in a potential window from 0.3 V$_{RHE}$ to -0.9 V$_{RHE}$ in 1 M KHCO$_3$. A reference spectrum obtained at 0.05 V$_{RHE}$ in 1 M KOH is subtracted. (c) Chronoamperogram and calculated faradaic efficiencies for CO$_2$ electro-reduction products after 1 h potentiostatic electrolysis at -0.3, -0.6, and -0.9 V$_{RHE}$.

C-bound COOH$_{ad}$ was found to convert into CO$_{ad}$ on Pt at potentials < 0.05 V$_{RHE}$; it eventually poisoned the surface and suppressed further reduction steps as well as the HCOO(K) formation via the one-step direct hydrogenation of physisorbed CO$_2$. A broad feature was observed at *ca.* 1550 cm$^{-1}$ in the SEIRA spectra at 0.3 V$_{RHE}$, which could be deconvoluted into two bands at 1543 and 1495 cm$^{-1}$ (Figure 3c). These peaks could be assigned to the C=O stretching mode of hydrogenated CO$_2$ adsorbates (COOH$_{ad}$) and (H)CO$_{3,ad}$, respectively [75]. When accounting for the presence of the solvent at an implicit level, the frequency computed for the C=O stretching of COOH$_{ad}$ is 1649 cm$^{-1}$. This value shifts down to 1620 cm$^{-1}$ when accounting for a hydrogen-bonded water molecule (Figures 3b and S6, Table S2). Following an opposite trend, the computed C=O-stretch frequency for $HCO_3$ in implicit water is 1355 cm$^{-1}$, but it shifts up to 1376 cm$^{-1}$ when explicitly accounting for a water molecule in the first coordination shell. At 0.2 V$_{RHE}$, the COOH$_{ad}$ band (1543 cm$^{-1}$) was red-shifted and reduced in intensity; the increase in intensity of two distinct bands at 1977 and 1762 cm$^{-1}$ could be assigned to the C≡O stretching mode of the linear and bridge form of CO$_{ad}$, respectively[76,77]. In addition, the CO$_{ad}$-related bands were present at potentials higher than the onset of the reduction current (*ca.* 0.2 V$_{RHE}$) in LSV (Figure 3a, inset), where the measured C≡O stretching frequencies (*ca.* 1975 cm$^{-1}$ for CO$_{ad,L}$ and 1760 cm$^{-1}$ for CO$_{ad,B}$) agree well with the values (*ca.* 1970 cm$^{-1}$ for CO$_{ad,L}$ and *ca.* 1755 cm$^{-1}$ for CO$_{ad,B}$) recently reported at similar potentials [78]. These values, however, are slightly redshifted (~50 cm$^{-1}$) as compared to literature values obtained under CO-saturated KOH conditions on Pt(111)[79]. This discrepancy could be explained through the presence of coadsorbed OH[80,81,82,83], which might be produced through the following conversion reaction mechanism from COOH$_{ad}$ to CO$_{ad}$: COOH$_{ad}$ → CO$_{ad}$ + OH$_{ad}$ (instead of the theoretically predicted COOH$_{ad}$ + H$_{ad}$ → CO$_{ad}$ + H$_2$O(g) [84,85]). Coadsorbed H$_2$O$_{ad}$/OH$_{ad}$ species, in fact, can be held responsible for the redshift of the C≡O stretching mode of CO$_{ad}$[75]. The predicted frequencies for linear- and bridge-CO$_{ad}$ in implicit solvent are 2074 and 1831 cm$^{-1}$, respectively. These values agree well with the experimental peak positions if accounting for the redshift due to the OH adsorption. The SEIRA spectra were dominated by the CO$_{ad,L}$ (1977 cm$^{-1}$) and CO$_{ad,B}$ (1762 cm$^{-1}$) bands at potentials below 0.05 V$_{RHE}$, suggesting that the surface was mainly covered by CO adsorbates in this potential range. No change in the wavenumber of the CO$_{ad,L}$ (1977 cm$^{-1}$) and CO$_{ad,B}$ (1762 cm$^{-1}$) bands between 0.05 and -0.9 V$_{RHE}$ indicates that the Stark tuning shift[47] was counterbalanced by another process: One possibility is the displacement of the coadsorbed OH$_{ad}$ by H$_{ad}$[75], which is supported by the observation of H$_2$ in the potential range from -0.3 to -0.9 $V_{RHE}$. The latter is confirmed by on-line gas chromatography (GC) (Figure 3c), which is indicative of HER occurring on the CO-covered Pt surface in this potential region. Finally, at potentials lower than -0.5 $V_{RHE}$, a weak and broad vibrational band appeared at *ca.* 1390 cm$^{-1}$, which could be attributed to the symmetric O-C-O stretching mode of an interfacial non-adsorbed HCOOK (or HCOO$^-$) located at the vicinity of the surface (referred as HCOO$^-_{int}$),[38] due to the absence of a corresponding SEIRAS active mode of the bridge-bonded HCOO$_{ad}$ on Pt (*ca.* 1320 cm$^{-1}$) [86] [87] and suppression of adsorbed HCOO at pH > 6 [88]. Here we propose a similar reaction mechanism to the one reported for Cu (100)[89] to account for the presence of HCOO$^-_{int}$ at <-0.5 $V_{RHE}$ on the CO-covered Pt surface, which includes the one-step direct hydrogenation of physisorbed $CO_2$ reacting with $H_{ad}$ ($CO_2$(gas) + $H_{ad}$ + e$^-$ → HCOO$^-$)[89]. As the reaction mechanism proposed does not involve adsorbates with highly stable Pt-C bonds, the HCOO$^-$ (or HCOOK) produced can leave the surface without passivating it. In contrast to previous findings[4], we could not detect HCOO$^-$ (or HCOOK) in the NMR measurements (Figure 3d), whereas HCOOK (or HCOO$^-$) produced on Pt and associated with low availability of $H_{ad}$ (due to the fast kinetics of HER and $CO_{ad}$ poisoning) might be below the detection limit of this study.

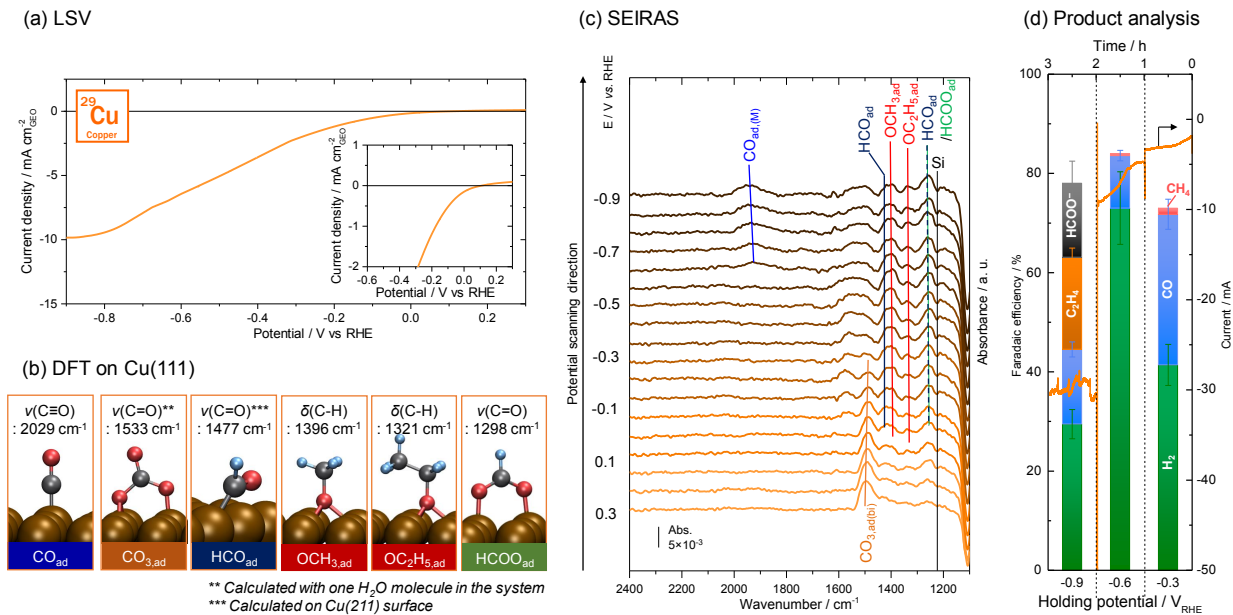

Figure 4 *In situ* ATR-SEIRAS measurement and product analysis on a Cu electrode in 1 M $KHCO_3$ (pH=6.8). (a) Linear sweep voltammogram in 1 M $KHCO_3$ at a scan rate of 10 mV $s^{-1}$. Inset shows the magnified linear sweep voltammogram from 0.3 to -0.6 $V_{RHE}$. (b) Calculated infrared-active vibrational frequencies of possible adsorbed intermediates on Cu(111) and Cu(211) and their schematic structures. Cu, C, O, and H atoms are depicted as brown, dark-gray, red and light blue spheres respectively. (c) *In situ* ATR-SEIRA spectra obtained during linear sweep voltammetry in a potential window from 0.3 $V_{RHE}$ to -0.9 $V_{RHE}$ in 1 M $KHCO_3$. A reference spectrum obtained at 0.05 $V_{RHE}$ in 1 M KOH is subtracted. (c) Chronoamperogram and calculated faradaic efficiencies for $CO_2$ electro-reduction products after 1 h potentiostatic electrolysis at -0.3, -0.6, and -0.9 $V_{RHE}$.

*In situ* SEIRAS measurements of Cu with optimal carbon monoxide binding relative to Pt and Au[3] [21] [22] [23] and strong binding with oxygen[20] revealed that bidentate $CO_{3,ad}$ was mainly reduced to O-bound intermediates, including $OCH_{3,ad}$ (1390 $cm^{-1}$), $OCH_2CH_{3,ad}$ (1340 $cm^{-1}$), and bidentate $HCOO_{ad}$ (1240 $cm^{-1}$), from 0.1 $V_{RHE}$ to -0.3 $V_{RHE}$ on the Cu surface (Figure 4). The distinct band at 1495 $cm^{-1}$ at 0.3 $V_{RHE}$ could be assigned to the C=O stretching mode of the bidentate form of $CO_{3,ad}$ (bound to the surface through two oxygen atoms)[38]. The assignment is supported by the DFT calculations, showing that bidentate $CO_{3,ad}$ is the only stable carbonate on the two flat Cu surfaces considered (Cu(100) and Cu(111)). The calculated C=O stretching frequencies on Cu(100) and Cu(111), with an explicit water molecule in the first-solvation shell, are 1510 $cm^{-1}$ and 1533 cm$^{-1}$ respectively, which agree well with the experimentally obtained value of 1495 cm$^{-1}$ (Figures 4b and S5, Table S3). Bands related to C-bound adsorbates such as COOH$_{ad}$ (~1550 cm$^{-1}$) or CO$_{ad}$ (~2000 cm$^{-1}$) were not observed on Cu at 0.3 V$_{RHE}$, indicating that Cu surfaces prefer M-O binding to M-C binding due to strong affinity with oxygen,[20] in contrast to what was observed for Pt and Au surfaces under similar conditions. The intensity of the bidentate CO$_{3,ad}$ band at 1495 cm$^{-1}$ was reduced with potentials lower than 0.1 V$_{RHE}$ and the band completely disappeared at -0.2 V$_{RHE}$, which was accompanied by the growth of three peaks at 1390, 1340, and 1240 cm$^{-1}$ (Figure 4c). Based on the DFT results, these observed peaks can be attributed to the CH bending of O-bound OCH$_{3,ad}$ (1392-1396 cm$^{-1}$), OCH$_2$CH$_{3,ad}$ (1320-1321 cm$^{-1}$), and to the symmetric C=O stretching of bidentate HCOO$_{ad}$ (1298-1306 cm$^{-1}$), respectively, where the values in parenthesis correspond to the computed frequency ranges for the three surface terminations considered (more details in Table S1 and Figure S5). These findings confirm earlier theoretical predictions[25] according to which OCH$_{3,ad}$ is an intermediate for the formation of CH$_4$, and trace amounts of CH$_4$ were detected by on-line GC at -0.3 V$_{RHE}$ (Figure 4d).

Further decreasing the potential below -0.3 V$_{RHE}$ first stabilizes C-bound species on the Cu surface, as supported by changes in the SEIRA spectra. A band at ~1420 cm$^{-1}$ appeared at potentials lower than -0.3 V$_{RHE}$, which could be attributed to the C=O stretching of hydrogenated CO (HCO$_{ad}$), having a C-bound configuration on a step-edge site, on the basis of our calculated frequency (1477 cm$^{-1}$). The computed frequency for the O-C-H bending of HCO$_{ad}$ is at 1246 cm$^{-1}$, which would be overshadowed by the bidentate HCOO$_{ad}$ peak (~1240 cm$^{-1}$). The presence of such hydrogenated CO (HCO$_{ad}$) is key to trigger the C-C dimerization reaction, which is known to represent a crucial step for the ethylene production from CO$_2$ and CO[12,90,91,92]. Upon decreasing the applied potential to -0.6 V$_{RHE}$, a broad feature appeared at ~1900 cm$^{-1}$, which could be assigned to the C≡O stretching mode of linear $CO_{ad}$ (the DFT calculations gave 2020-2043 cm$^{-1}$ in Table S1) and accounting for Stark-tuning associated with low potentials applied. The appearance of the $CO_{ad}$ band is consistent with a decrease in the Faradaic efficiency for CO formation (Figure 4d). Therefore, SEIRA spectra show that the Cu surface is able to stabilize intermediates not only with a O-bound configuration (e.g. $OCH_{3,ad}$, $OCH_2CH_{3,ad}$, and $HCOO_{ad}$) but also with a C-bound configuration, such as $CO_{ad}$ and $HCO_{ad}$ below -0.6 $V_{RHE}$. Considering that the $HCO_{ad}$ band (*ca.* 1420 cm$^{-1}$) appeared at -0.3 $V_{RHE}$ and below, while the $CO_{ad}$ bands emerged only below -0.7 $V_{RHE}$ just before the detection of ethylene by on-line GC at potential -0.9 $V_{RHE}$ (Figure 4d), we propose that the stabilization of $CO_{ad}$ is required to trigger the C-C bond formation for ethylene formation[12] [90] [91] [92]. This proposed reaction mechanism for ethylene formation thus involves the C-C coupling between $CO_{ad}$ and $(H)CO_{ad}$ intermediates ($CO_{ad}$ + $(H)CO_{ad}$ => $OC\text{-}CO(H)_{ad}$), followed by subsequent hydrogenation ($OC\text{-}COH_{ad}$ + $4H_2O$ + $6e^-$ => $H_2C\text{=}CH_{ad}$ + $6OH^-$) and complete with cleavage of Cu-C bond ($H_2C\text{=}CH_{ad}$ + $H_2O$ + $e^-$ => $CH_2\text{=}CH_2$ + $OH^-$), which is similar to the proposed CO reduction pathway on Cu to produce ethylene [93]. Although $OCH_2CH_{3,ad}$ has also been suggested as a potential precursor for ethylene formation,[94] the intensity of the corresponding peak at 1340 cm$^{-1}$ did not decrease even at -0.9 $V_{RHE}$, suggesting that $OCH_2CH_{3,ad}$ is unlikely to contribute to the ethylene formation observed in this study.

In addition to ethylene, NMR- and GC-detected products at -0.9 $V_{RHE}$ included formate, CO, and $H_2$ (Figure 4d), in agreement with previous work[12] [95] [94] [96]. Although Cu was shown to form a considerable amount of formate (HCOO-) at -0.9 $V_{RHE}$ (Figure 4d), no clear decrease in intensity of the $HCOO_{ad}$ band at 1240 cm$^{-1}$ was observed in the SEIRA spectra. Our reasoning is as follows: 1) the only SEIRAS active vibrational mode for $HCOO_{ad}$ is at 1240 cm$^{-1}$ and would overlap with one of the $HCO_{ad}$ bands and 2) similar to the Pt surface, the formation of the HCOO$^-$ can proceed via the one-step direct hydrogenation of physisorbed $CO_2$ ($CO_2$(gas) + $H_{ad}$ + $e^-$ → $HCOO^-$)[89] and does not require $HCOO_{ad}$ intermediates. In contrast, no alcohols (e.g. methanol and ethanol) were detected by NMR at the investigated potentials, which is consistent with the monotonic growth of the $OCH_{3,ad}$ and $OCH_2CH_{3,ad}$ peaks at 1390 $cm^{-1}$ and 1340 $cm^{-1}$, respectively. This observation suggests that those intermediates are not able to desorb from the surface due to the strong binding with the Cu surface[20], which can be a major limiting step for alcohol formation on Cu.

In summary, C-bound intermediates (e.g. $HCO_{ad}$ and $CO_{ad}$) can be stabilized on Cu surfaces at relatively low potentials (<-0.3 $V_{RHE}$), which is required for forming products including CO and $C_2H_4$. On the other hand, $CH_4$ was the only product we were able to detect from O-bound intermediates ($OCH_{3,ad}$) in this study. $CH_4$ is formed through breaking of the O-C bond instead of the Cu-O bond, while breaking Cu-O bond in O-bound intermediates (e.g. $CO_{3,ad}$, $OCH_{3,ad}$, and $OC_2H_{5,ad}$) is believed to be a required step to produce alcohols on Cu.

3.3. Discussion of the CORR mechanisms

Following the detailed experimental and theoretical analysis of the reaction intermediates as a function of potential, we propose the metal-dependent CORR mechanisms illustrated in Scheme 1.

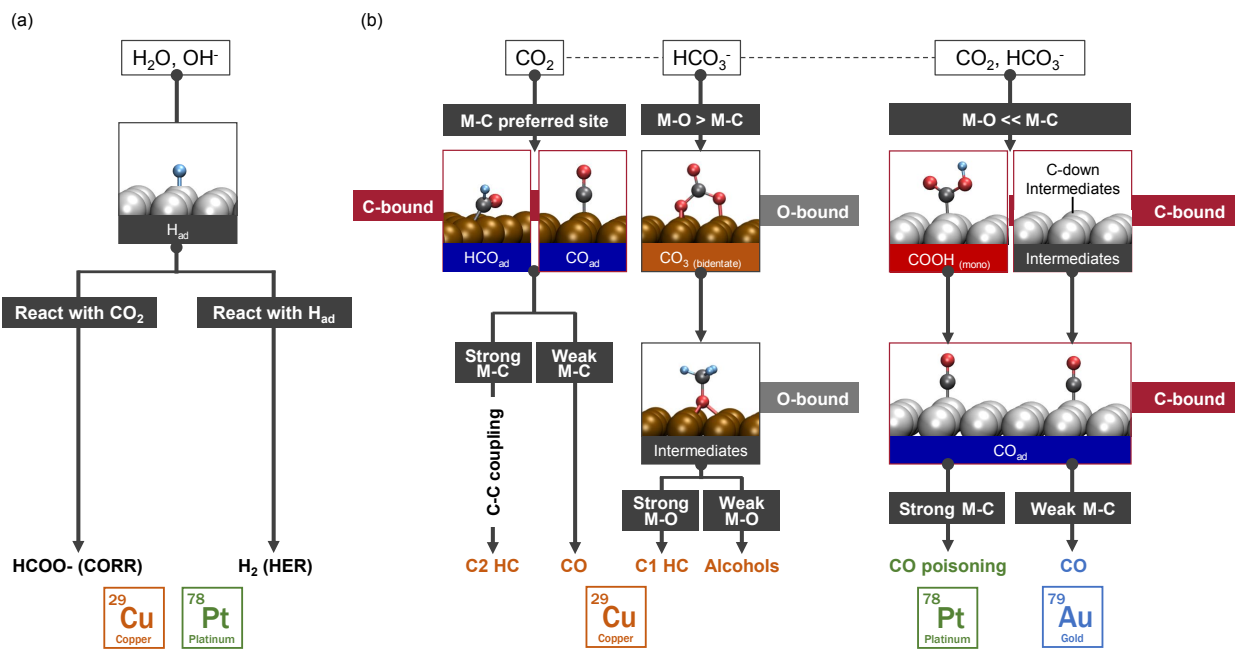

Scheme 1 Summary of the proposed reaction mechanisms for the CORR on Au, Cu, and Pt electrodes during the potential scan between 0.05 V$_{RHE}$ and 1.4 V$_{RHE}$ in 1 M KHCO$_3$. Proposed reaction pathway for (a) formate and hydrogen evolution and (b) hydrocarbon, CO, alcohol formation reaction. *M*, Cu, C, O, and H atoms are depicted as gray, brown, dark-gray, red and light blue spheres, respectively (*M*=Au, Pt, and Cu).

Considering that no change in the bands related to adsorbed formate (HCOO$_{ad}$) accompanies the appearance of formate species in solution, we deduce that the major reaction mechanism for formate formation on Pt and Cu shown in Scheme 1a is the one-step direct hydrogenation via physisorbed CO$_2$ reacting with H$_{ad}$. Formate formation on Pt and Cu has been thought to compete with the CO formation and proceed through hydrogenation of the carbon of CO$_{2,ad}$, either via Eley−Rideal (ER, CO$_{2,ad}$ + H$_2$O + $e^-$ => HCOO$_{ad}$ + OH$^-$)[2][96][89] and/or Langmuir−Hinshelwood (LH, CO$_{2,ad}$ + H$_{ad}$ => HCOO$_{ad}$) mechanisms[2][96][89]. However, corresponding intermediates (CO$_{2,ad}$) were absent in our SEIRAS measurement, in agreement with having the Gibbs free energies for the ER pathway (1.12 eV)[89] and LH pathway (0.99 eV) [89] higher than the proposed pathway of one-step direct hydrogenation of physisorbed CO$_2$ reacting with H$_{ad}$ (CO$_2$(gas) + H$_{ad}$ + e$^-$ → HCOO$^-$ , 0.37 eV)[89]. We propose that this formate mechanism is distinct from the reaction pathway forming CO (on Pt and Cu) and hydrocarbons (on Cu) in this study, which competes with the hydrogen evolution reaction. A surface with optimal H-binding, such as Pt, prefers the HER while Cu with slightly weaker H-binding than Pt leans more towards formate evolution. The reaction of $H_{ad}$ with $CO_2$ to produce formate is infrequent on Au, the metal with weakest H-binding[20], probably because of the low stability (short life) as well as low coverage (low-frequency factor) of $H_{ad}$. The selectivity of this pathway can be controlled by the H-binding capability of the catalysts, since $H_{ad}$ is the only adsorbate involved in the direct hydrogenation process.

The proposed reaction mechanisms shown in Scheme 1b can be explained by the M-C and M-O affinity trends for the three metals. On Pt and Au, the intermediates observed at low overpotentials always have a C-bound configuration, such as $COOH_{ad}$ (1543 cm$^{-1}$) on Pt and $CO_{ad}$ (*ca.* 2100 cm$^{-1}$) on Au. These results support previous experimental[37,97] and theoretical studies[85,98] on Pt and Au suggesting C-bound $COOH_{ad}$ (or $CO_{2,ad}{}^{\delta -}$) is hydrogenated to form $CO_{ad}$ and $H_2O$. The fate of the C-bound intermediates on the two surfaces can be determined by the M-C binding strength. With strong M-C binding such as Pt-C, $COOH_{ad}$ on Pt is further reduced to $CO_{ad,L}$ (1977 cm$^{-1}$) and $CO_{ad,B}$ (1762 cm$^{-1}$) at potentials below 0.2 $V_{RHE}$, which eventually poison the surface. Therefore, the difficulty in converting C-bound intermediates into O-bound species limits the production kinetics of C1 hydrocarbons ($CH_4$) and alcohols ($CH_3OH$)[94]. On the other hand, on Au, $CO_{ad}$ is gradually displaced by $H_{ad}$ and/or $K_{ad}$ below -0.05 $V_{RHE}$, indicating the weak interaction between CO and the Au surface. Bidentate (bi)carbonate adsorbates are not stable on Au even at 0.65 $V_{RHE}$ (Figure S4), suggesting a considerably weak interaction between Au and O-bound intermediates as well.

For the Cu surface, we propose that the initial intermediate configuration (O-bound vs C-bound) determines the final product; O-bound intermediates are generally responsible for C1 hydrocarbons and alcohols production while C-bound intermediates are required to produce C2 hydrocarbons and CO. The initial intermediate have been suggested to be C-bound intermediates such as $CO_{2,ad}^{\delta-}$, which can be further reduced to $CO_{ad}$ and $CHO_{ad}$ via subsequent hydrogenation[25] [94] [99]. Also $CHO_{ad}$ is shown by DFT studies to be the key intermediates to form both C1 and C2 hydrocarbons through hydrogenation, with subsequent C-O cleavage [20] [25] [94] [99] and C-C bond formation[94] [100], respectively. Although proposed mechanisms from DFT include transformation of the intermediate configuration (for C1 hydrocarbon case, C-bound $*CH_xO_{ad}$ to O-bound $CH_x*O_{ad}$ ($x$ = 1 or 2) [25] [94]), we could not detect any SEIRAS bands corresponding to O-bound $CH_xO_{ad}$ ($x$ = 1 or 2) intermediates. On the basis of the SEIRAS observation, we have proposed an alternate pathway that initiates with the Cu-(bi)carbonate interaction to form O-bound $CO_{3,ad}$, which could lead to the formation of C1 hydrocarbons and alcohols. O-bound $CO_{3,ad}$ intermediates (1495 cm$^{-1}$) are dominant on Cu at low overpotential (~0 $V_{RHE}$), which is in agreement with the binding energy trend for M-O(H) (Cu<Pt<Au, also see Figure S4) [20] [21]. At potentials of -0.3 $V_{RHE}$ or lower, C-bound intermediates such as $HCO_{ad}$ (1420, 1243 cm$^{-1}$) and $CO_{ad}$ (1900 cm$^{-1}$) become sufficiently stable on the Cu surface. These intermediates are likely formed via the adsorption of $CO_2$ with subsequent reduction ($CO_2 + H_2O + 2e^- => CO_{ad} + 2OH^-$ or $CO_2 + 2H_2O + 3e^- => HCO_{ad} + 3OH^-$). The co-existence of C-bound and O-bound intermediates is possible probably because of the difference in binding energy on each surface termination, of which there are many on polycrystalline catalysts (e.g. the M-C binding energy is lower on Cu(100) than on Cu(111) [91] [101] [102]). It is important to stress the fact that the initial O-bound intermediates ($CO_{3,ad}$) can be obtained from (bi)carbonate in the solution, whereas C-bound intermediates ($HCO_{ad}$ and $CO_{ad}$) are likely produced via reductive adsorption of $CO_2$. The proposed mechanism thus suggests the importance of the local pH in the vicinity of the electrode surface, which is expected to control the $CO_2$/(bi)carbonate ratio, on the product selectivity of Cu. Such hypothesis would rationalize the reported cation effect on product selectivity for Cu, where increasing the cation size ($Rb^+$ and $Cs^+$) increases the buffering capability ($pK_a$ of hydrolysis) and decreases the pH near the surface, thus leading to a significant increase in the product ratio of C2 hydrocarbon (ethylene) to C1 hydrocarbon (methane) [103]. The absence of C-bound intermediates in SEIRAS in the low overpotential region (~0 $V_{RHE}$) indicates that Cu generally prefers having Cu-O rather than Cu-C bonds, which further supports the notion that O-bound intermediates ($OCH_{3,ad}$ and $OC_2H_{5,ad}$) bind strongly on the Cu surface once they form below -0.3 $V_{RHE}$. Therefore, alcohols are among the most challenging products to form via the proposed reaction mechanism, because of the necessary trade-off between the binding of the key O-bound intermediates ($OCH_{3,ad}$ and $OC_2H_{5,ad}$) and their ability to desorb from the catalyst surface.

## 5. Conclusions

In this work, we examine three different metal catalysts (Pt, Cu, and Au), having different affinity for C-bound and O-bound adsorbates, which are generally considered as key reaction intermediates of the CORR. We apply *in situ* SEIRAS coupled with DFT calculations in order to understand the surface (electro)chemistry as well as to establish a catalyst design strategy to improve product selectivity for the CORR.

Pt has been found to be inactive for CORR and only produced $H_2$ from HER. The surface is covered by $CO_{ad}$ adsorbates below 0 $V_{RHE}$, which behave as poisonous species once they are formed on the surface. For Au, $CO_{ad}$ adsorbates are detected as the only CORR product, in addition to $H_2$ formation from HER. No higher-order C1 and C2 fuels have been observed. On the other hand, CO, formate, methane, and ethylene are produced during CORR on Cu. These products are accompanied by the spectroscopic observation of O-bound $CO_{3,ad}$, $OCH_{3,ad}$ and $OC_2H_{5,ad}$, as well as C-bound $(H)CO_{ad}$.

The proposed reaction mechanisms can be rationalized in terms of catalyst preferences for C-bound and O-bound intermediates. The initial intermediates that are formed on the surfaces are controlled by the balance between the affinity for M-C and M-O binding, and different outcomes are observed depending on the surface catalytic properties: on Pt, $CO_{ad}$, which is obtained from $COOH_{ad}$, is so stable that ends up poisoning the surface; on Au, $CO_{ad}$ leaves the surface as CO gas due to the weak Au-CO interaction; on Cu, the final products of CORR are determined by the configuration of the initial adsorbates, C-bound ($HCO_{ad}$) and O-bound ($CO_{3,ad}$), which can be obtained from $CO_2$ and (bi)carbonate in the electrolyte, respectively. The O-bound $CO_{3,ad}$ intermediates can be subsequently reduced into O-bound intermediates ($OCH_{3,ad}$), which are suggested to be responsible for C1 hydrocarbon (methane) production, while the C-bound intermediates (($H)CO_{ad}$) are responsible for C2-hydrocarbon (ethylene) and CO production. These results indicate a simple way of controlling the product selectivity by changing the pH in the vicinity of the surface and by tuning the electrolyte composition. The proposed reaction mechanism also reveals a dilemma in the tuning of the M-O binding to design alcohol-selective catalysts: stronger M-O binding would accelerate the formation of O-bound intermediates ($OCH_{3,ad}$ and $OC_2H_{5,ad}$), which are considered as important alcohol precursors; at the same time, these would suppress the intermediate hydrogenation and alcohol desorption. Thus, the requirements for an alcohol-selective catalyst cannot be fulfilled by simply designing the electronic structure of the catalyst to tune its adsorbate binding capability. High selectivity for the CORR should instead be achieved by tuning new knobs including, but not limited to, the non-covalent interactions between intermediates and ions in the solution as well as the interfacial water structure, that can control the electrolyte conditions at the vicinity of the electrode.

ASSOCIATED CONTENT

**Supporting Information Available:**

Si 2p and O1s spectra for Pt, Cu, and Au under 100 mTorr $CO_2$ and 10 mTorr $H_2O$ in AP-XPS. Representative C 1s spectrum collected at 490 eV incident photon energy for one of the annealing and sputtering cycles for Au, Cu, and Pt. *In situ* ATR-SEIRA spectra obtained chronoamperometry at 0.25, 0.35, 0.45, and 0.65 $V_{RHE}$ for 10 mins in CO-saturated 1 M KOH and 1 M $K_2CO_3$ (pH=11.5). Computed vibrational frequencies ($cm^{-1}$) and simulated SEIRAS spectra for the selected adsorbates on the three Cu surface termination considered, as well as on Pt(111) and Au(111). This material is available free of charge via the Internet at http://pubs.acs.org.

AUTHOR INFORMATION

**Corresponding Author**

*Corresponding author. Tel.: +1-617-253-2259(Y. S.-H.), +81-(0)836-85-9285(Y. K.)

E-mail address: shaohorn@mit.edu (Y. S.-H.), yuktym@yamaguchi-u.ac.jp (Y. K.)

**Author Contributions**

The manuscript was written with contributions from all authors. All authors have approved the final version of the manuscript.

ACKNOWLEDGMENTS

Y.K. thanks the Japan Gateway Kyoto University Top Global Program for travel funds to Massachusetts Institute of Technology (MIT). Work at MIT was supported by Eni S.p.A.. Y. K. was supported in part by Grant-in-Aid for JSPS Research Fellow. This research used resources of the National Energy Research Scientific Computing Center, a DOE Office of Science User Facility supported by the Office of Science of the U.S. Department of Energy under Contract No. DE-AC02-05CH11231. This work also used resources of the Extreme Science and Engineering Discovery Environment (XSEDE) [104], which is supported by National Science Foundation grant number ACI-1548562. This project has received funding from the European Union's Horizon 2020 research and innovation programme under grant agreements No. 665667 and No. 798532.